\documentclass[fleqn,10pt]{wlscirep}
\usepackage{braket}

\title{Boron Triangular Kagome Lattice with Half-Metallic Ferromagnetism}

\author[1,*]{Sunghyun Kim}
\author[1]{W. H. Han}
\author[2]{In-Ho Lee}
\author[1]{K. J. Chang}
\affil[1]{Department of Physics, Korea Advanced Institute of Science and Technology, Daejeon 34141, Korea}
\affil[2]{Korea Research Institute of Standards and Science, Daejeon 34113, Korea}

\affil[*]{kimsunghyun@kaist.ac.kr}

\affil[+]{these authors contributed equally to this work}

\begin{abstract}
Based on the first-principles evolutionary materials design, we report a stable boron Kagome lattice composed of triangles in triangles on a two-dimensional sheet.
The Kagome lattice can be synthesized on a silver substrate, with selecting Mg atoms as guest atoms.
While the isolated Kagome lattice is slightly twisted without strain, it turns into an ideal triangular Kagome lattice under tensile strain.
In the triangular Kagome lattice, we find the exotic electronic properties, such as topologically non-trivial flat band near the Fermi energy and half-metallic ferromagnetism, and predict the quantum anomalous Hall effect in the presence of spin-orbit coupling.
\end{abstract}
\begin{document}

\flushbottom
\maketitle

\section*{Introduction}
Recently, two-dimensional (2D) materials have attracted much attention because of their unusual characteristics, such as Dirac fermions, topological states, and valley polarization\cite{Geim2005,Qian2014,Geim2013}.
The discovery of graphene, a monolayer of carbon atoms in the honeycomb lattice\cite{Geim2005}, has resulted in the synthesis of different 2D materials and various graphene analogues, such as hexagonal BN, transition metal dichalcogenides, silicene, germanene, stanene, and phosphorene\cite{Geim2013,Novoselov2016,Balendhran2015}.
Elemental boron exhibits a variety of allotropes with structural complexity\cite{Albert2009,Oganov2009}, but a 2D honeycomb lattice is inherently prohibited because boron lacks one valence electron compared to its adjacent carbon in the periodic table. 
A flat triangular boron lattice has been reported to be unstable to a buckled shape due to excessive electrons occupying the antibonding states\cite{Tang2007}.
Since the surplus of electrons can be balanced by introducing hexagonal holes in the triangular lattice, more stable 2D forms composed of triangular and hexagonal motifs have been proposed, including the B $\alpha$-sheet and its analogues\cite{Tang2007, Penev2012}.

A Kagome lattice also consists of triangular and hexagonal motifs in a network of corner-sharing triangles.
Many exotic phenomena have been predicted for the Kagome lattice, such as frustrated magnetic ordering\cite{Balents2010,Nisoli2013}, ferromagnetism\cite{Mielke1992,Tanaka2003}, and topologically non-trivial states\cite{Ohgushi2000,Guo2009,Tang2011}.
However, the experimental realization of the Kagome lattice is confined to the Kagome layers of pyrochlore oxides\cite{Gardner2010}, the self-assembled metal-organic molecules on a substrate\cite{Mao2009}, and the cold atoms of an optical Kagome lattice\cite{Jo2012}.
A B$_{3}$ Kagome lattice, in which a triangular motif is made of three B atoms, is known to be dynamically unstable due to the lack of electrons\cite{Xie2015}.
Theoretical calculations indicate that, in the case of 2D metal-B systems such as MoB$_{4}$\cite{Xie2014}, TiB$_{2}$\cite{Zhang2014}, FeB$_{6}$\cite{Zhang2016}, FeB$_{2}$\cite{Zhang2016a}, and MnB$_{6}$\cite{Li2016}, triangular or honeycomb B networks can be stabilized by the electron transfer from metal ions to B networks and the interaction between metal and B layers.
Similarly, a bilayer form of MgB$_6$ sandwiching the Mg layer between two B$_{3}$ Kagome layers was found to be stable\cite{Xie2015}.
Despite a number of theoretical attempts to predict 2D boron allotropes, only a few 2D boron sheets have been synthesized on metal substrates, such as a quasi-2D layer of $\gamma$-B$_{28}$\cite{Tai2015}, a 2D triangular sheet (generally referred to as borophene)\cite{Mannix2015}, and borophene with stripe-patterned vacancies\cite{Feng2016}.
Given the structural diversity of metal-B systems, a proper choice of substrate and metal elements as guest atoms can open the way to realizing stable 2D boron Kagome sheets that have not yet been discovered. 

In this work, we perform an evolutionary crystal structure search for 2D boron phases with the Mg atoms as guest atoms on a silver substrate.
We find a new 2D boron sheet consisting of triangular B networks and the Mg atoms embedded in large hexagonal voids.
The boron sheet separated from the substrate forms a twisted Kagome lattice and turns into an ideal triangular Kagome lattice under tensile strain, accompanied with a metal-to-half-metal transition.
The ferromagnetism of the triangular Kagome lattice is characterized by a nearly flat band at the Fermi level, which is topologically non-trivial and thus induces the quantum anomalous Hall effect in the presence of spin-orbit coupling.

\section*{Results}

\subsection*{Crystal structure search}
First, we explored two-dimensional Mg-B allotropes with low energies on the substrate 
by using an {\it ab initio} evolutionary crystal structure search method, as implemented in the AMADEUS code\cite{Lee2016}.
Distinct configurations were generated under the constraint of layer group symmetry, with the number of configurations setting to 20 in the population size of global optimization.
For each configuration, the energy minimization was performed by using the generalized gradient approximation of Perdew, Burke, and Ernzerhof (PBE)\cite{Perdew1996} for the exchange-correlation potential and the projector augmented wave pseudopotentials\cite{Blochl1994}, as implemented in the VASP code\cite{Kresse1999}.
The details of calculations are given in Method.

\begin{figure}
\includegraphics[width=\columnwidth]{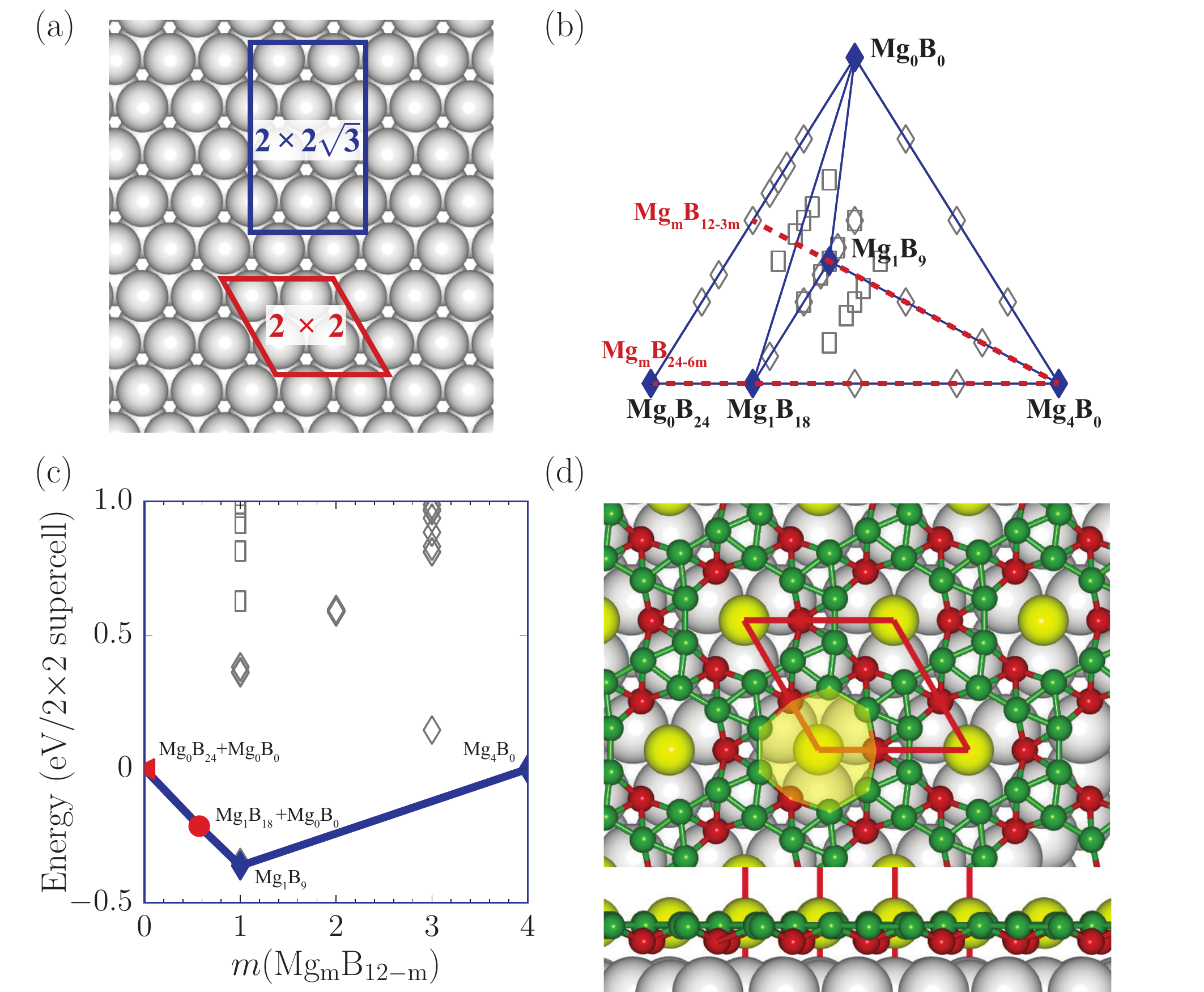}
\caption{\label{fig:convexhull}
\textbf{Phase diagram of Mg-B systems on Ag(111).}
(a) Rhombic $2 \times 2$ and rectangular $2 \times 2\sqrt{3}$ supercells on the Ag(111) surface. (b) A phase diagram of 2D Mg-B systems on the substrate.
In the $2 \times 2$ (rhombuses) and $2 \times 2\sqrt{3}$ (rectangles) supercells, filled and open symbols stand for the stable and metastable states, respectively, and red dashed lines represent the B monolayer and bilayer coverages.
(c) The convex hull diagram and (d) the lowest energy structure of Mg$_{1}$B$_{9}$ for the monolayer coverage are drawn. Yellow, green, and gray balls represent the Mg, B, and Ag atoms, respectively, and red balls denote the B atoms sharing large triangles in the Kagome lattice. }
\end{figure}

\begin{figure}
\includegraphics[width=\columnwidth]{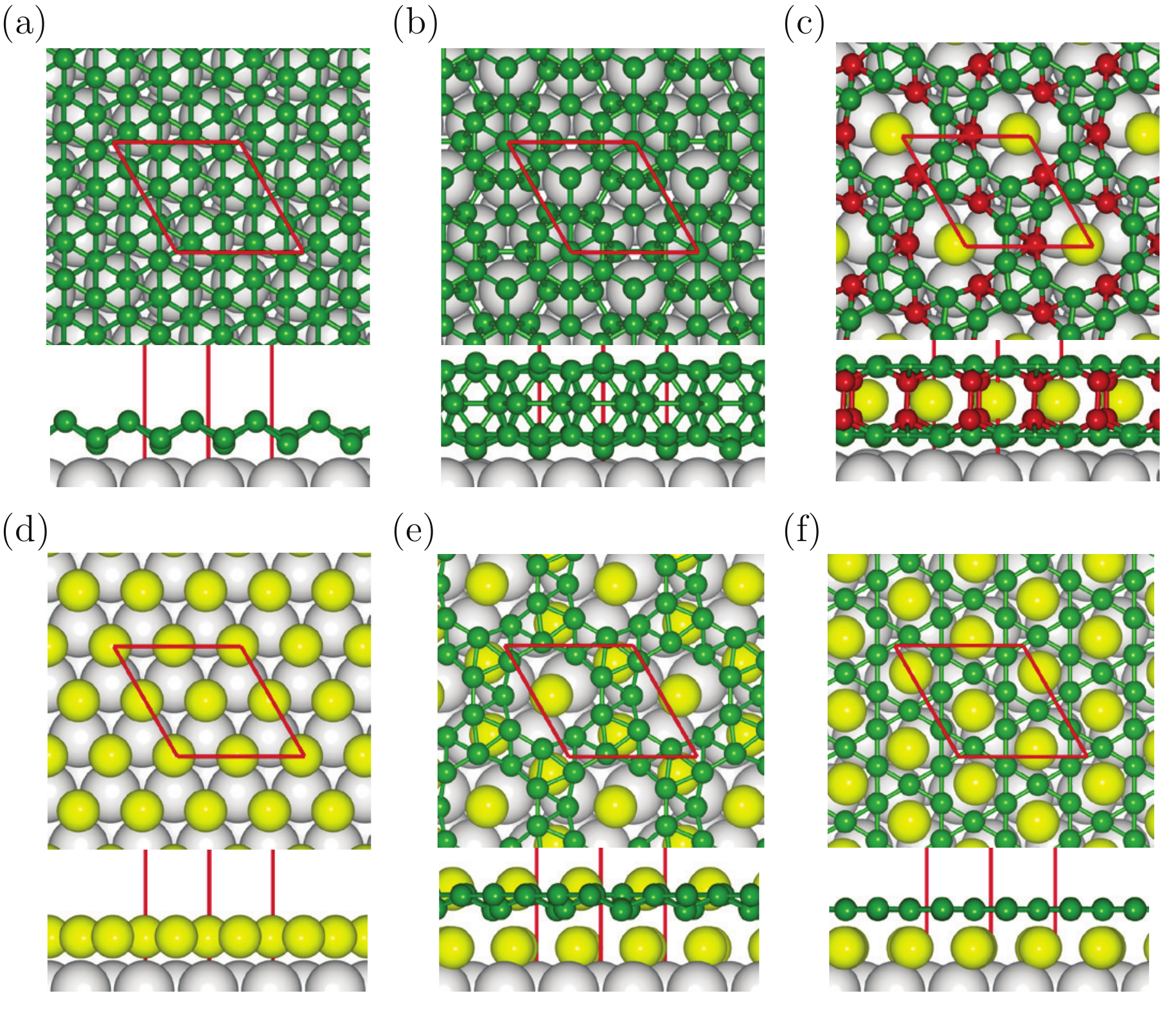}
\caption{\label{fig:configurations}
\textbf{Atomic structures of B-Mg systems on Ag(111).}
The atomic structures of (a) Mg$_0$B$_{12}$ (referred to as borophene), (b) Mg$_0$B$_{24}$, (c) Mg$_4$B$_{0}$, and (d) Mg$_0$B$_{18}$ are drawn. The Mg$_0$B$_{24}$ structure consists of B$_{20}$ rectified hexagonal bipyramids connected by four B atoms. The atomic structures of  are shown for (e) the metastable and (f) most stable configurations. Yellow, green, and gray balls represent the Mg, B, and Ag atoms, respectively. Red rhombuses represent the $2\times2$ supercell on the Ag(111) surface.}
\end{figure}

We considered monolayer and bilayer coverages of the B atoms with the Mg atoms as guest atoms for two lateral supercells on the Ag(111) surface, rhombic $2 \times 2$ and rectangular $2 \times 2\sqrt{3}$ (Figure \ref{fig:convexhull}a).
If a 2D triangular sheet, which has been synthesized on a Ag substrate\cite{Mannix2015}, is taken as the most stable structure for the monolayer coverage, the B atoms can be deposited up to 12 in the $2 \times 2$ supercell (Figure \ref{fig:configurations}a).
Since the Mg atoms prefer to occupy the hollow sites on the Ag(111) surface, they can be deposited up to 4 for the monolayer coverage, and three B atoms per Mg atom are depleted.
In the bilayer coverage, the Mg atoms pop into the empty space between the two B layers, keeping the same maximum coverage of 4, and each Mg atom depletes six B atoms.
Considering such combinations of Mg and B, 2D Mg-B phases can be represented as Mg$_{m}$B$_{n}$ with $0 \leq m \leq 4$ and $0 \leq n \leq 12-3m$ ($0 \leq n \leq 24-6m$) for the B monolayer (bilayer) coverage in the $2 \times 2$ supercell.

We searched for the lowest-energy configurations for Mg$_{m}$B$_{n}$ systems and compared their formation energies defined as,
\begin{equation}\label{eq:energy_convexhull}
\begin{split}
E(\mathrm{Mg}_{m}\mathrm{B}_{n}) = E_{tot}(\mathrm{Mg}_{m}\mathrm{B}_{n}) -x E_{tot}(\mathrm{Mg}_{4}\mathrm{B}_{0}) 
-y E_{tot}(\mathrm{Mg}_{0}\mathrm{B}_{24}) -z E_{tot}(\mathrm{Mg}_{0}\mathrm{B}_{0}),
\end{split}
\end{equation}
where $E_{tot}(A)$ is the total energy of the system $A$ and $x = m/4$, $y = n/24$, and $z = 1-x-y$.
Here, Mg$_{4}$B$_{0}$ and Mg$_{0}$B$_{24}$ correspond to the Mg monolayer and the B bilayer on the Ag surface, respectively, whereas Mg$_{0}$B$_{0}$ denotes the bare Ag substrate (Figures \ref{fig:configurations}b and \ref{fig:configurations}c).
The bilayer form of Mg$_{0}$B$_{24}$ is energetically more favorable than the Mg$_{0}$B$_{12}$ monolayer on the Ag substrate. 
  Thus, the chemical potentials of Mg and B on the substrate are defined as, 
  $\mu_\mathrm{Mg}$ = [$E_{tot}$(Mg$_{4}$B$_{0}$) $-$ $E_{tot}$(Mg$_{0}$B$_{0}$)]/4 and 
  $\mu_\mathrm{B}$ = [$E_{tot}$(Mg$_{0}$B$_{24}$) $-$ $E_{tot}$(Mg$_{0}$B$_{0}$)]/24.
The phase diagram of Mg$_{m}$B$_{n}$ is shown in Figure~\ref{fig:convexhull}b, and the configurations on the convex hull are energetically stable against the decomposition into other phases.
Among Mg$_{m}$B$_{n}$ systems, we find that Mg$_{1}$B$_{9}$ lies on the convex hull for the monolayer coverage (Figure~\ref{fig:convexhull}c).
The stability of Mg$_{1}$B$_{9}$ was confirmed through the tests for both the $2 \times 2$ and $2 \times 2\sqrt{3}$ supercells.
The Mg$_{1}$B$_{9}$ allotrope consists of nine B atoms in slightly buckled triangular networks and one Mg atom occupying a large hole in the unit cell (Figure~\ref{fig:convexhull}d).
As the boron coverage increases, we obtained a bilayer form of Mg$_{1}$B$_{18}$ on the convex hull, in which the Mg atoms are sandwiched between the two B layers, lying in between the empty holes.
In the Mg$_{1}$B$_{18}$ allotrope, each B layer has the same triangular network as that of Mg$_{1}$B$_{9}$, however, its buckling is enhanced due to interlayer interactions (Figure \ref{fig:configurations}d).

\subsection*{Atomic structure and stability}
\begin{figure}
\includegraphics[width=\columnwidth]{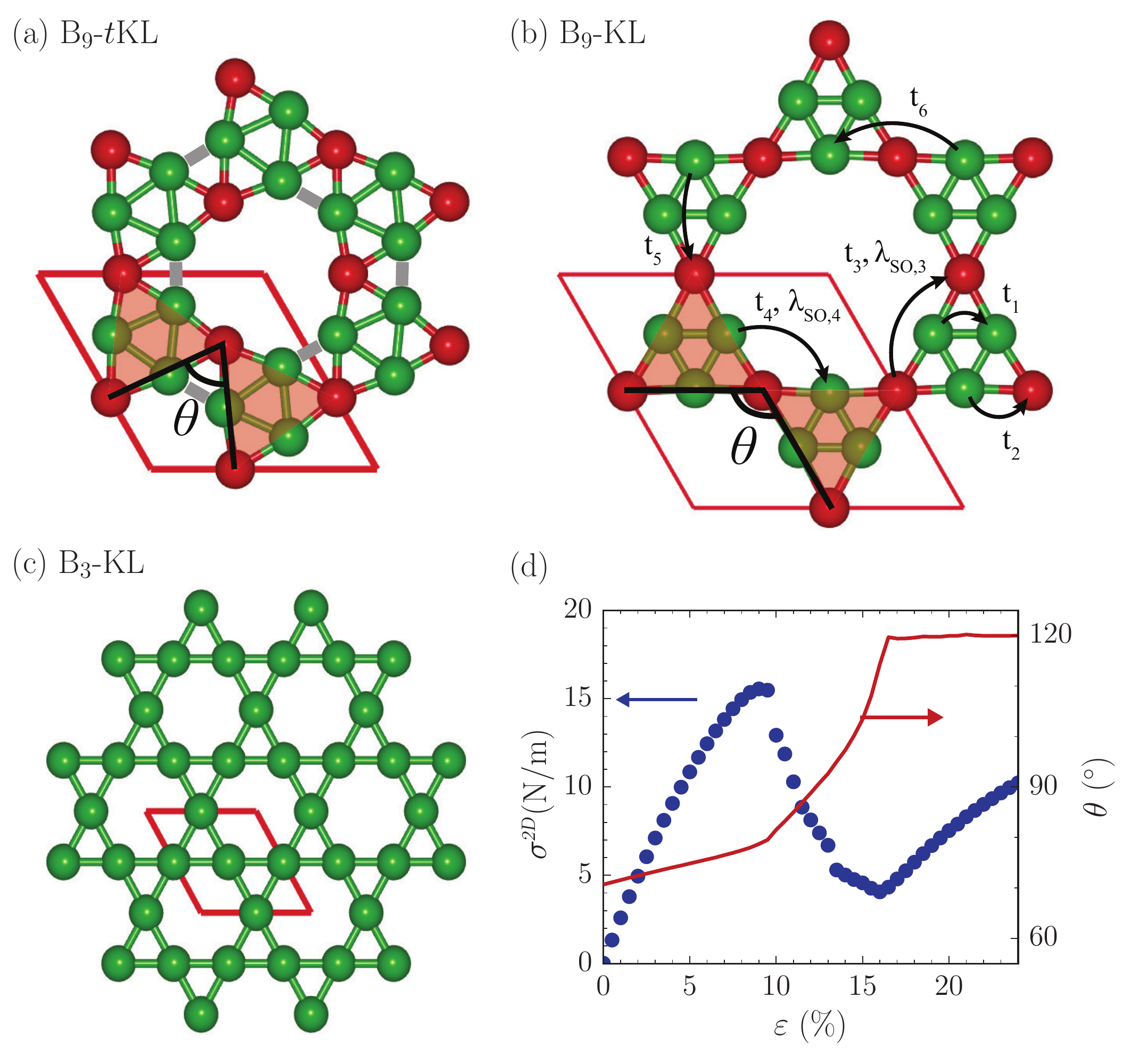}
\caption{\label{fig:structure}
\textbf{Strain effect on atomic structure.}
Atomic structures for (a) B$_{9}$-$t$KL, (b) B$_{9}$-KL, and (c) B$_{3}$-KL. Red balls denote the B atoms sharing large triangles in B$_{9}$-$t$KL and B$_{9}$-KL. In (a), gray lines represent the bonds between large triangles, which are broken in B$_{9}$-KL.
(d) The 2D stress ($\sigma^{2D}$) and the angle between two large triangular units ($\theta$) are plotted as a function of biaxial strain. }
\end{figure}

A 2D boron Kagome lattice can be obtained from Mg$_{1}$B$_{9}$ 
by exfoliating the Mg-B sheet from the substrate and removing the guest atoms.
The isolation of a 2D boron sheet can be made by using various exfoliation techniques\cite{Nicolosi2013,Tao2015}.
In a free-standing Mg$_{1}$B$_{9}$ sheet, the Mg ions bind weakly to the boron networks with a smaller binding energy of about 0.7 eV/Mg (Supplementary Table S1), compared with other magnesium borides which were suggested as the potential cathode materials for Mg-ion batteries\cite{Zhao2011}.
This result indicates that the Mg ions in the Mg$_{1}$B$_{9}$ sheet can be dissolved in conventional electrolytes used for Mg batteries\cite{Saha2014,Zhang2016b}.
In the optimized Mg-free B$_{9}$ sheet, called a twisted Kagome lattice (denoted as B$_{9}$-$t$KL),
three B atoms are depleted per empty hole, and each void is surrounded by six large triangles (Figure \ref{fig:structure}a).
As biaxial tensile strain ($\varepsilon$) is applied, the voids are enlarged,
and the bonds connecting large triangular units are subsequently broken, 
resulting in an ideal triangular Kagome lattice (denoted as B$_{9}$-KL)\cite{Norman1990,Mekata1998}, as shown in Figure \ref{fig:structure}b.
We find that the B$_{9}$-$t$KL and B$_{9}$-KL sheets are perfectly flat,
while the remaining Mg ions cause buckling or twisting of the B networks due to the charge transfer.
Both B$_{9}$-$t$KL and B$_{9}$-KL belong to the general Kagome system with the subnet 2, in which each triangle of the Kagome arrangement contains a stack of four triangles\cite{Ziff2009}.
The lattice parameters, plane groups, and Wyckoff positions of B$_{9}$-$t$KL and B$_{9}$-KL are given in Table~\ref{table:structure}.

To estimate the critical strain ($\varepsilon_c$) for the transition from B$_{9}$-$t$KL to B$_{9}$-KL, 
we calculated the 2D biaxial stress ($\sigma^{2D}$) as a function of strain.
In the stress-strain curve, we find two stress drops at $\varepsilon$ = 9.5\% and 13\% (Figure \ref{fig:structure}d).
The prominent drop at $\varepsilon$ = 9.5\% is accompanied with bond-breaking relaxations between the large triangular units, whereas the weak drop at $\varepsilon$ = 13\% is related to a transition to the ferromagnetic state (which will be discussed shortly).
When the bonds between the edge B atoms of large triangles are broken, the coordination number of the edge B atoms is reduced from 5 to 4.
Then, a charge transfer of 2.5 electrons occurs from six edge atoms to three corner atoms within the unit cell (Table \ref{table:structure}).
As strain increases above 9.5\%, the angle between the large triangular units ($\theta$) increases rapidly and reaches 120$^{\circ}$ at the critical strain of 16.5$\%$, where B$_{9}$-KL is formed (Figure \ref{fig:structure}d).
For strain above 16.5\%, B$_{9}$-KL experiences only elastic deformation and maintains the angle of $\theta$ = 120$^{\circ}$.
For strain up to 24\%, overall the calculated stress is below 16 $\rm{N/m}$, lying in the stress range accessible by using an atomic force microscope, as demonstrated for various 2D materials, such as graphene and transition-metal dichalcogenides\cite{Lee2008,Bertolazzi2011}.

\begin{figure}[ht]
\includegraphics[width=\columnwidth]{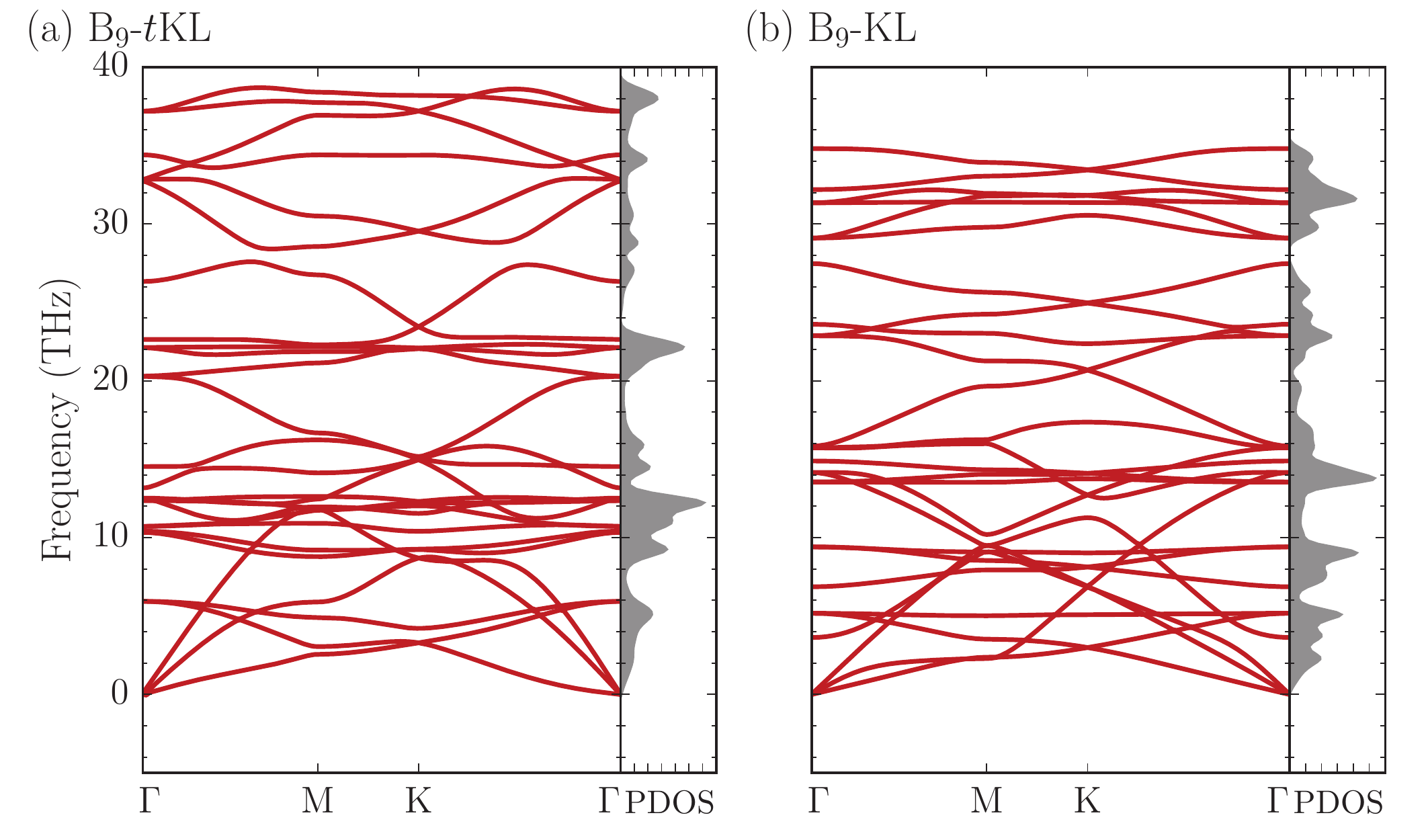}
\caption{\label{fig:phonon}
\textbf{Phonon spectra.}
The calculated phonon spectra are drawn for (a) B$_{9}$-$t$KL (without strain) and (b) B$_{9}$-KL (under 18\% strain), with the phonon density of states (PDOS in arbitrary units) on the right panels.}
\end{figure}

The stability of B$_{9}$-$t$KL and B$_{9}$-KL was verified by calculating the full phonon spectra and 
performing first-principles molecular dynamics simulations at high temperatures (Figure\ref{fig:phonon} and Supplementary Figure S1). 
Among 2D boron sheets composed of triangular and hexagonal motifs, a B$_{3}$ Kagome lattice (denoted as B$_{3}$-KL) also has a network of corner-sharing triangles, similar to B$_{9}$-$t$KL and B$_{9}$-KL.
In B$_{3}$-KL, however, each triangle of the Kagome arrangement is made of three B atoms, and a single B atom is depleted in each hexagonal hole (Figure \ref{fig:structure}c).
It is known that B$_{3}$-KL is dynamically unstable due to one-electron deficiency to fully occupy the bonding states.
On the Ag substrate, we find a Mg$_{3}$B$_{9}$ structure consisting of the B$_{3}$ Kagome lattice and the Mg atoms located underneath the hexagonal holes (Figure \ref{fig:configurations}e), similar to MgB$_6$\cite{Xie2015}.
However, this allotrope is energetically less stable by 1.64 eV per $2 \times 2$ cell than the combined structure of Mg$_{1}$B$_{9}$ with two additional Mg atoms sandwiched between the Mg$_{1}$B$_{9}$ layer and substrate (Figure \ref{fig:configurations}f).

The valence band of B$_{3}$-KL consists of five bonding states: two three-center $\sigma$-bonding states, two $\sigma$-bonding states captured in hexagonal holes, and one delocalized $\pi$-bonding state.
In B$_{9}$-KL, six edge atoms in two adjacent large triangular motifs form one six-center $\sigma$-bonding state, replacing for two three-center $\sigma$-bonding states of B$_{3}$-KL.
While the number of the $\sigma$-bonding states is reduced by one, two $\pi$-bonding states are fully occupied, in contrast to B$_{3}$-KL, and the $\pi$-non-bonding state is half-filled due to an exchange splitting (Supplementary Figure S2).
Meanwhile, one $\sigma$-bonding state is further reduced due to the multi-center bonds between the edge atoms of large triangles in B$_{9}$-$t$KL.
Although one $\sigma$-antibonding state is half-filled, two $\pi$-bonding and one $\pi$-non-bonding states are fully filled.
Therefore, it is inferred that the occupation of the delocalized $\pi$-orbital states plays a crucial role in stabilizing the flat forms of both B$_{9}$-$t$KL and B$_{9}$-KL, similar to the cases of B $\alpha$-sheet\cite{Galeev2011} and B clusters\cite{Zhai2003}.

\subsection*{Electronic structure}
\begin{figure}
\includegraphics[width=\columnwidth]{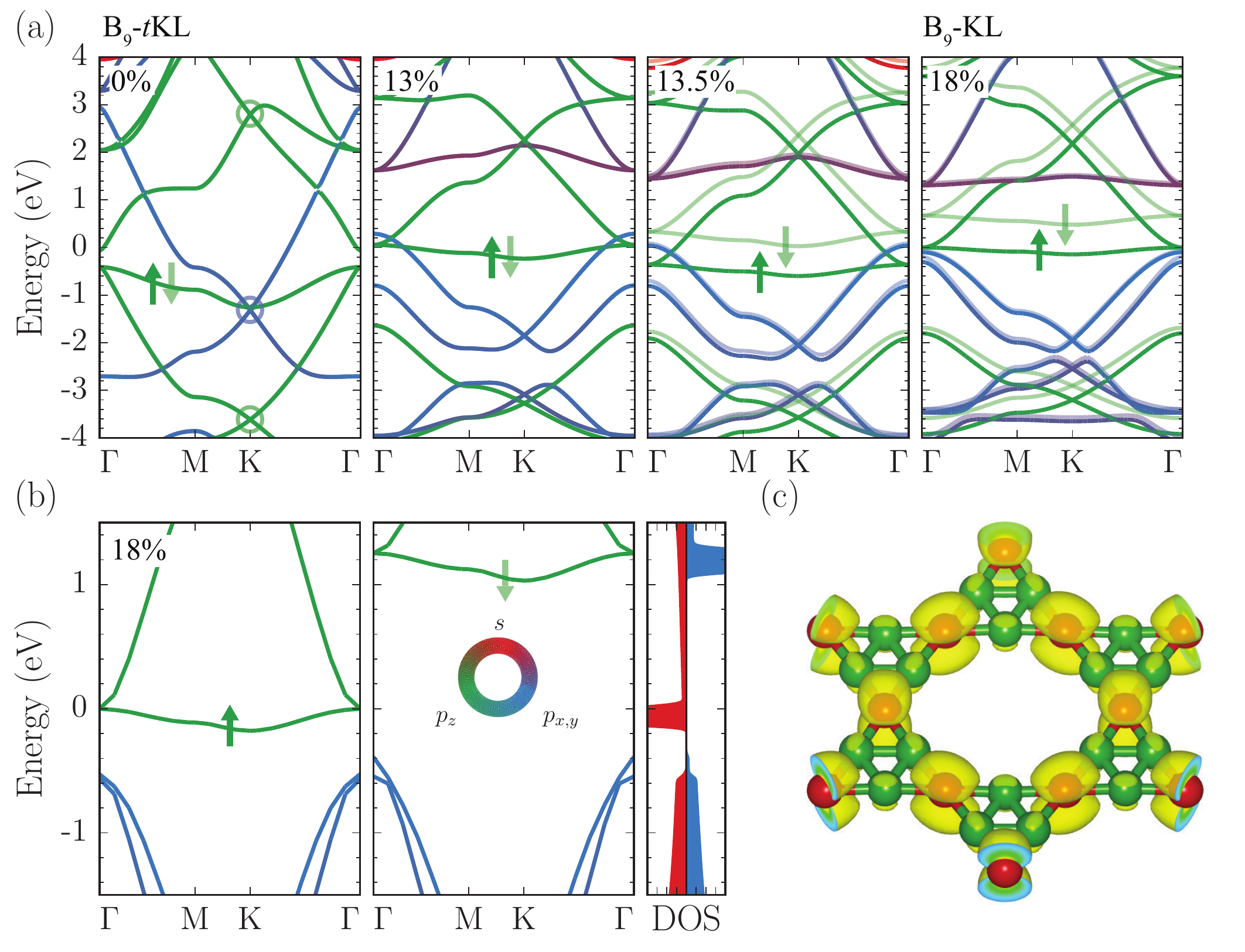}
\caption{\label{fig:bandstructure} 
\textbf{Band structure evolution under strain.}
(a) The PBE band structures of B$_{9}$-$t$KL and B$_{9}$-KL under strain.
Circles indicate three Dirac-like bands, and the bands are color-coded according to their orbital characteristics: green for $p_{z}$, blue for $p_{x}$ and $p_{y}$, and red for $s$ orbitals. The spin-down bands are shaded in lighter colors.
(b) The HSE06 band structure of B$_{9}$-KL for spin-up and spin-down electrons, with the spin-resolved density of states on the right panel. (c) The contour plot of the charge density for the flat band near the Fermi level at the $\rm{K}$ point in B$_{9}$-KL. }
\end{figure}

We find that B$_{9}$-KL exhibits a half-metallic band structure, whereas B$_{9}$-$t$KL is metallic but nonmagnetic (Figure~\ref{fig:bandstructure}a).
In B$_{9}$-$t$KL, a less dispersive band appears at about $-1.0$ eV below the Fermi level, and three Dirac-like bands are formed at the K point: one characterized by the $p_{x}$ and $p_{y}$ orbitals in the valence band and two by the $p_{z}$ orbital in both the valence and conduction bands.
While strain changes the positions of three Dirac-like bands, a noticeable effect is that the less dispersive band moves upward and becomes flattened.
Thus, the density of states at the Fermi level singificantly increases, causing the Stoner instability\cite{Stoner1947}.
For strain above 13\%, the exchange splitting occurs for the flat band, resulting in a half-filled band structure and thereby a ferromagnetic transition, similar to the flat-band ferromagnetism\cite{Mielke1992,Tanaka2003}.
With the hybrid functional of Heyd, Scuseria, and Ernzerhof (HSE06) for the exchange-correlation potential\cite{Heyd2003}, for B$_{9}$-KL, we find that the band width of the flat band is 175 meV and the band gap of spin-down electrons is 1.42 eV (Figure~\ref{fig:bandstructure}b).

\begin{figure}
\includegraphics[width=\columnwidth]{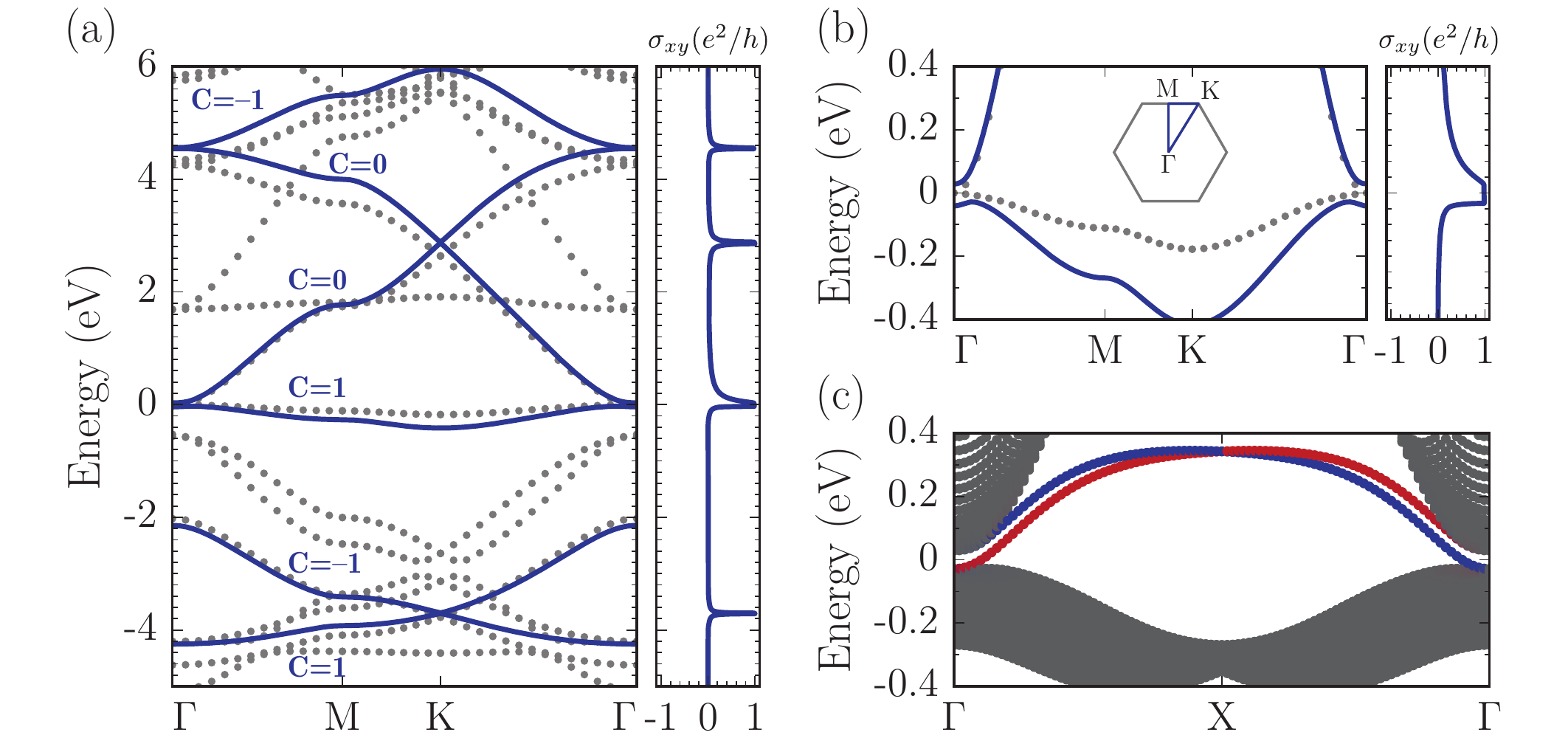}
\caption{\label{fig:tb_model} 
\textbf{Tight-binding modeling.}
(a) The spin-up band structure of B$_{9}$-KL under 18\% strain and (b) its magnified view near the Fermi level along high symmetry lines, with the Hall conductances on the right panels. Gray dots stand for the HSE06 band structure and 
blue curves represent the band structure derived from the effective tight-binding model with including the SOC.
(c) The band structure of the B$_{9}$-KL ribbon with a width of 50 unit cells.
Blue and red curves stand for the chiral bands associated with the different edges. }
\end{figure}

To understand the origin of the flat band in B$_{9}$-KL, we analyzed the wave function associated with the flat band, which is mainly derived from the localized $p_{z}$ orbitals at the corner B atoms of triangles (Figure~\ref{fig:bandstructure}c). 
By using maximally localized Wannier functions (MLWFs)\cite{Mostofi2008}, we derived an effective tight-binding Hamiltonian, with the hopping parameters, $t_{1}=-2.31$ eV, $t_{2}=-2.10$ eV, $t_{3}=-0.08$ eV, $t_{4}=0.29$ eV, $t_{5}=0.52$ eV, and $t_{6}=0.39$ eV (see Supplementary Method and Figure \ref{fig:structure}b).
Nine MLWFs are sufficient enough to represent the bonding characteristics between the $p_{z}$ orbitals
and well reproduces the HSE06 band structure near the Fermi level, as illustrated in Figure \ref{fig:tb_model}a.
If only the nearest-neighbor hopping terms are considered, a completely flat band is formed at the Fermi level 
because of the destructive interference of the hopping terms to adjacent hexagonal holes (Supplementary Figure S3).

In the effective tight-binding Hamiltonian, we introduced the spin-orbit coupling (SOC), with the parameters $\lambda_\mathrm{SO,3}$ = 0.004 eV and $\lambda_\mathrm{SO,4}$ = 0.015 eV, which correspond to 5\% of the second nearest-neighbor hopping terms, $t_{3}$ and $t_{4}$, respectively.
Although the SOC is small for B systems, it could be enhanced by various methods,
such as hydrogenation\cite{Balakrishnan2013}, introduction of transition metal adatoms\cite{Weeks2011}, and substrate proximity effects\cite{Calleja2015}, which have been used for graphene.
With including the SOC, we find that the band gap of 57 meV opens at the $\Gamma$ point and 
the flat band has the non-trivial Chern number of $C=1$ (Figure \ref{fig:tb_model}a).
This result implies that the quantum anomalous Hall effect could be realized in B$_{9}$-KL, provided that the SOC opens the band gap.
In fact, the quantization of Hall conductance is found in B$_{9}$-KL with the SOC (Figures \ref{fig:tb_model}a and \ref{fig:tb_model}b), which is the hallmark of the quantum anomalous Hall effect.
In addition, the appearance of chiral edge states inside the bulk band gap is confirmed in a one-dimensional ribbon of B$_{9}$-KL (Figure \ref{fig:tb_model}c).

\section*{Conclusion}
In conclusion, we have predicted the ideal planar shape of a triangular B$_{9}$ Kagome lattice, which contains triangles in triangles.
Based on the {\it ab initio} evolutionary crystal structure search, we propose that the B$_{9}$ Kagome lattice can be synthesized on the Ag(111) surface, with selecting the Mg atoms as guest atoms.
The triangular B$_{9}$ Kagome lattice offers the exotic electronic characteristics, such as flat band at the Fermi level and half-metallic ferromagnetism, providing opportunities for spintronics applications.
Because of the non-trivial band topology of the flat band, the quantum anomalous Hall effect can be realized in the presence of spin-orbit coupling.
\\
\section*{Methods}
\subsection*{Density functional calculation}
We employed the generalized gradient approximation of Perdew, Burke, and Ernzerhof (PBE) \cite{Perdew1996} for the exchange-correlation potential and the projector augmented wave pseudopotentials \cite{Blochl1994}, as implemented in the VASP code \cite{Kresse1999}.
In addition, van der Waals forces were taken into account to describe more accurately interlayer interactions \cite{Grimme2006}.
In a slab geometry, a vacuum region larger than 15 {\AA} was inserted, with a dipole correction \cite{Neugebauer1992}, ensuring prohibiting interactions between adjacent supercells, and only the topmost layer of the silver substrate was relaxed.
The wave functions were expanded in plane waves up to an energy cutoff of 300 eV and the Monkhorst-Pack mesh \cite{Monkhorst1976} with a grid spacing of $2\pi \times 0.03$ {\AA}$^{-1}$ was used for Brillouin zone integration.
The atomic coordinates were optimized until the residual forces were less than 0.02 eV/{\AA}.
For the free-standing B${_9}$-$t$KL, the in-plane lattice vectors were relaxed until stress was below 0.5 kbar.
At the final stage of optimization, we used a higher energy cutoff of 400 eV.

\bibliography{bib}

\section*{Acknowledgements}
This work was supported by Samsung Science and Technology Foundation under Grant No. SSTF-BA1401-08.

\section*{Author contributions statement}
S.K., W.H.H., I.-H.L, and K.J.C. contributed equally to this manuscript. K.J.C. conceived the work and designed the research strategy. S.K. performed theoretical calculations and S.K. and W.H.H. did data analysis. All authors discussed the results and co-wrote the manuscript.

\section*{Additional information}
\textbf{Competing financial interests} The authors declare no competing financial interests.

\clearpage
\begin{table}
\centering
\caption{\label{table:structure}
The calculated lattice constants ($a$), plane groups, Wyckoff positions, and Bader charges ($q$) are compared for B$_{9}$-$t$KL (zero strain) and B$_{9}$-KL (18\% strain).}
\begin{tabular}{cccccc}
\hline 
                 &   $a$ (\AA)& Plane group & Wyckoff position & $q$ ($e$) & \\  \hline
  B$_{9}$-$t$KL  & 5.88      & $p31m$ (No. 15) & 3$c$ (0.368, 0)         & 2.33 & \\
                 &           &                & 6$d$ (0.314, 0.479)    & 3.33 & \\
  B$_{9}$-KL     & 6.93      & $p6mm$ (No. 17) & 3$c$ (0.5, 0)         & 3.16 & \\
                 &           &                & 6$f$ (0.256, 0.512) & 2.91 & \\
\hline
\end{tabular}
\end{table}
\clearpage

\end{document}